\documentclass[prx, preprintnumbers, amsmath, amssymb, superscriptaddress, twocolumn]{revtex4-1}
\usepackage{graphicx, color, dcolumn, natbib, bm, amssymb, amsmath,color}
\graphicspath{{fig/}}
\usepackage[final=true]{hyperref}
\usepackage[normalem]{ulem}

\begin{document}
\title{Magnon-polaron excitations in the noncollinear antiferromagnet Mn$_3$Ge}
\author{A. S. Sukhanov}\thanks{alexandr.sukhanov@cpfs.mpg.de}
\affiliation{Max Planck Institute for Chemical Physics of Solids, D-01187 Dresden, Germany}
\affiliation{Institut f{\"u}r Festk{\"o}rper- und Materialphysik, Technische Universit{\"a}t Dresden, D-01069 Dresden, Germany}
\author{M. S. Pavlovskii}
\affiliation{Kirensky Institute of Physics, Siberian Branch, Russian Academy of Sciences, Krasnoyarsk 660036, Russian Federation}
\author{Ph. Bourges}
\affiliation{Laboratoire L\'eon Brillouin, CEA-CNRS, Universit\'e Paris-Saclay, CEA Saclay, 91191 Gif-sur-Yvette, France}
\author{H. C. Walker}
\affiliation{ISIS Facility, STFC, Rutherford Appleton Laboratory,
Didcot, Oxfordshire OX11-0QX, United Kingdom}
\author{K. Manna}
\affiliation{Max Planck Institute for Chemical Physics of Solids, D-01187 Dresden, Germany}
\author{C. Felser}
\affiliation{Max Planck Institute for Chemical Physics of Solids, D-01187 Dresden, Germany}
\author{D. S. Inosov}
\affiliation{Institut f{\"u}r Festk{\"o}rper- und Materialphysik, Technische Universit{\"a}t Dresden, D-01069 Dresden, Germany}
\begin{abstract}

We present the detailed inelastic neutron scattering measurements of the noncollinear antiferromagnet Mn$_3$Ge. Time-of-flight and triple-axis spectroscopy experiments were conducted at the temperature of 6~K, well below the high magnetic ordering temperature of 370~K. The magnetic excitations have a 5-meV gap and display an anisotropic dispersive mode reaching $\simeq 90$~meV at the boundaries of the magnetic Brillouin zone. The spectrum at the zone center shows two additional excitations that demonstrate characteristics of both magnons and phonons. The \textit{ab initio} lattice-dynamics calculations show that these can be associated with the magnon-polaron modes resulting from the hybridization of the spin fluctuations and the low-energy optical phonons. The observed magnetoelastic coupling agrees with the previously found negative thermal expansion in this compound and resembles the features reported in the spectroscopic studies of other antiferromagnets with the similar noncollinear spin structures.

\end{abstract}

\maketitle

\section{Introduction}

Magnons and phonons, which are quanta of spin waves and crystal-lattice vibrations in ordered materials, respectively, are a central topic in many areas of solid state research. The excitations of the lattice and the magnetic degrees of freedom as separate subsystems were extensively studied experimentally and theoretically over recent decades and have been described both qualitatively and quantitatively in many crystalline materials. Nowadays, the research focus has mainly shifted to understanding phenomena related to the interplay between lattice vibrations and spin excitations. Magnons and phonons, when strongly coupled together, can result in hybridised collective spin-lattice modes (magnetoelastic modes) -- magnon-phonon excitations, also referred to as magnon-polarons~\cite{ref1,ref2,ref3}.

Among the materials recognized for their significant magnon-phonon interaction, yttrium iron garnet (YIG) stands out having been actively studied in spin pumping experiments~\cite{ref2}, in measurements of magnetic field-dependent spin Seebeck effect~\cite{ref3}, in experiments on magnon-to-phonon conversion~\cite{ref4}, and in calculation of the magnon-phonon interconversion~\cite{ref5}, which can be applied in spintronics~\cite{ref6,ref7}. The presence of magnon-phonon coupling in YIG was also directly observed by neutron spectroscopy~\cite{ref_extra_1}. Magnetoelastic coupling was predicted to cause anomalous features in the transport properties of magnetic insulators~\cite{ref1} and greatly affect the magnon damping~\cite{ref8}.

Besides the ferrimagnetic compound, YIG, antiferromagnets (AFMs) with noncollinear arrangements of the magnetic moments form a broad family of materials that demonstrate strong hybridizations of the magnon and phonon modes. As was pointed out by the authors of Refs.~\cite{ref9,ref10}, noncollinear magnetic order allows for a one-magnon term in the spin Hamiltonian due to the absence of a global spin quantization axis. This leads to a linear magnon-phonon coupling, which is zero for collinear magnets. In other words, noncollinear AFMs possess excited states in which the magnetoelastic coupling is the primary source of the spectral renormalization, whereas the magnon-magnon interactions only influence higher-order corrections to the excitation spectrum.

Inelastic neutron scattering (INS) gives direct access to the spin-spin correlation function and serves as an effective probe for magnetic excitations. Recently, there were a number of INS studies reporting various features in the excitation spectra of AFMs with a 120$^{\circ}$ magnetic structure. The previously studied materials include the hexagonal multiferroics YMnO$_3$~\cite{ref10,ref11,ref12,ref13,ref14}, HoMnO$_3$~\cite{ref12,ref15}, and the trigonal delafossites CuCrO$_2$~\cite{Philippe,ref17} and LiCrO$_2$~\cite{ref9}. In all the listed materials, the discrepancy between the expected pure-magnon spectrum and the actually observed quasiparticle dispersions were discussed in relation to the emerging magnon-phonon hybridization. In addition, it was also demonstrated that the strong magnetoelastic coupling in YMnO$_3$ cardinally changes its thermal conductivity~\cite{ref_extra_2}. Despite the apparent similarity of the magnetic and crystal structures of these compounds, the spin-lattice interaction influenced the excitation spectra in distinct ways. This fact suggests that the excited states in magnets with magnetoelastic coupling depend not only quantitatively but also qualitatively on the details of the underlying exchange interactions, magnetic anisotropies, and phonon frequencies of the bare (uncoupled) subsystems. Interestingly, no signatures of spectrum renormalization were found in Lu$_{0.6}$Sc$_{0.4}$FeO$_3$~\cite{ref16}, which is a compound analogous to YMnO$_3$.

\begin{figure}[t]
        \begin{minipage}{0.99\linewidth}
        \center{\includegraphics[width=1\linewidth]{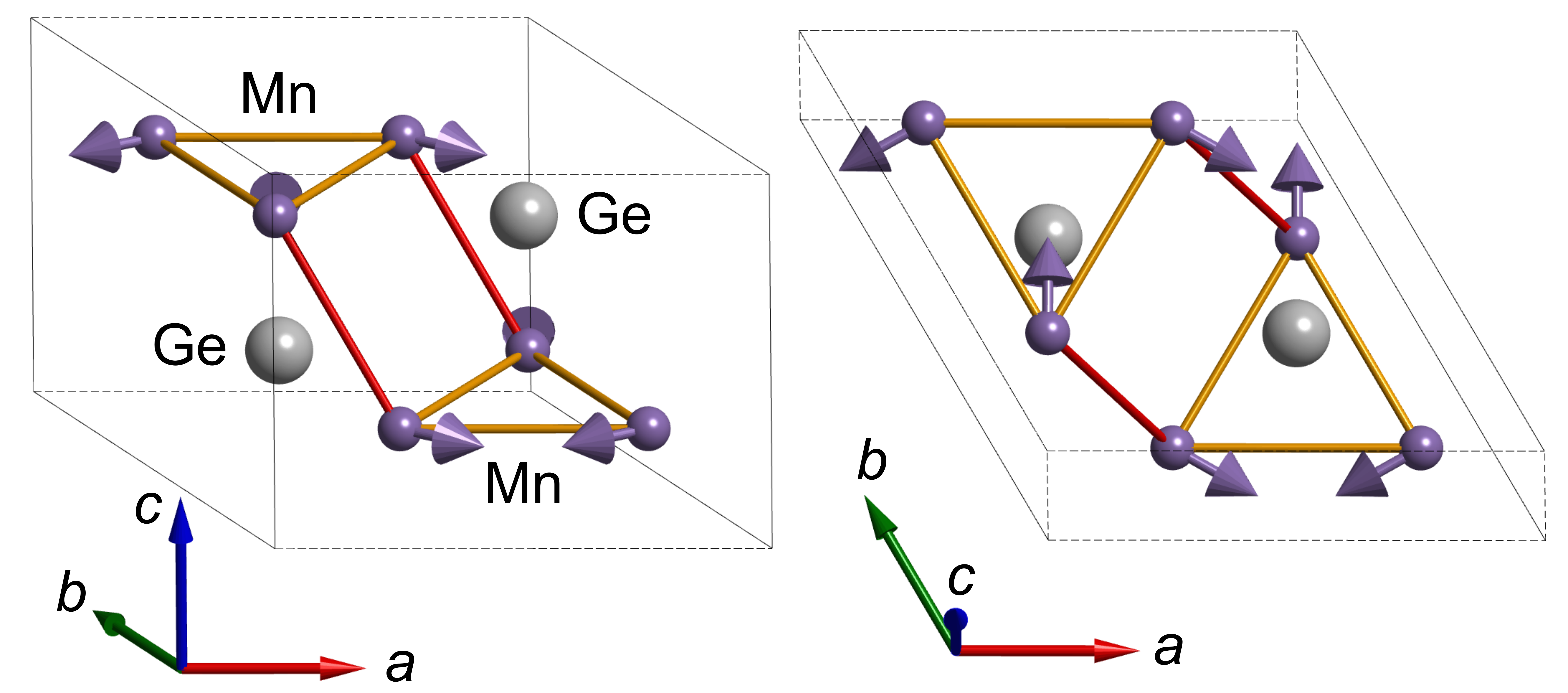}}
        \end{minipage}
        \caption{(color online). The chemical and magnetic unit cells of the noncollinear AFM Mn$_3$Ge shown in two different orientations. Arrows depict the magnetic moments, nearest and next-nearest Mn--Mn distances are shown by the red and orange lines, respectively.}
        \label{ris:fig0}
\end{figure}

In this paper we present the detailed INS study of another representative of the noncollinear AFMs with a 120$^{\circ}$ spin structure -- Mn$_3$Ge. The presence of a strong spin-lattice interaction in this compound was previously inferred from the observed negative thermal expansion~\cite{Sukhanov}. The temperature dependence of the lattice parameters in Mn$_3$Ge resembles the anomalies found in (Y,Lu)MnO$_3$~\cite{ref_extra_3} and in $R$MnO$_3$ ($R$~=~Ho, Yb, Sc)~\cite{ref12}. The metallic compound Mn$_3$Ge has the hexagonal symmetry $P6_3/mmc$ (space group~194). The crystal and magnetic structures of Mn$_3$Ge are sketched in Fig.~\ref{ris:fig0}. The chemical unit cell with lattice parameters $a = 5.39$~\AA~ and $c = 4.35$~\AA~ (at 300~K) contains two Ge atoms and six magnetic Mn atoms that form two kagome layers separated by a distance of $c/2$. The triangular magnetic order, which is characterized by the propagation vector $\textbf{k} = 0$, occurs below the high transition temperature $T_{\text N} \simeq 370$~K. The ordered magnetic moment of $\sim 2.5$~$\mu_{\text B}$ per Mn atom was extracted from previous powder neutron diffraction measurements at 100~K~\cite{Sukhanov}. In addition, polarized neutron diffraction measurements~\cite{weak1,weak2} indicated a presence of a weak in-plane ferromagnetic moment due to a slight distortion of an ideal 120$^{\circ}$ order, which amounts to $\sim 0.021$~$\mu_{\text B}$ per formula unit.

The isostructural compounds Mn$_3$Ge and Mn$_3$Sn recently attracted a broad attention due to discovery of a giant anomalous Hall effect (AHE), which was attributed to a symmetry properties of the noncollinear AFM structure of these materials~\cite{AHE1,AHE2,AHE3}. Unlike the well-known AHE in ferromagnets, which scales with the net magnetization of a material, the AHE in AFMs was long deemed to be impossible~\cite{AHE4,AHE5,AHE6,AHE7}. Moreover, recent studies of Mn$_3$Ge and Mn$_3$Sn indicated that these compounds might demonstrate the topological properties of their electronic band structure and belong to an unusual class of materials---Weyl semimetals~\cite{Weyl1,Weyl2,Weyl3}.

\section{Methods}

A large ($\sim$ 2.5 g) single crystal of Mn$_3$Ge was grown using the Bridgman--Stockbarger technique as described in Ref.~\cite{growth}. Inelastic neutron scattering (INS) measurements were conducted on the thermal direct time-of-flight (TOF) spectrometer Merlin~\cite{merlin1} located at the ISIS Neutron and Muon Source (Didcot, the UK) and the thermal triple-axis spectrometer (TAS) 2T1 at the LLB-Orph\'{e}e (CEA Saclay, France). Using x-ray backscattering Laue, the crystal was oriented in the horizontal $(HK0)$ plane for both experiments.

The optimization of both resolution and intensity in TOF measurements was achieved by setting the Fermi chopper frequency to 400~Hz. The incident neutron energy of 170~meV was chosen to cover a sufficiently large part of the 4D momentum-energy reciprocal space. We also used multirep mode~\cite{merlin2}, which allows two additional datasets with  $E_{\text{i}}$ = 59~meV and 29~meV to be simultaneously collected. This configuration resulted in an approximate energy resolution at the elastic line $\Delta E$ = 11, 2.8, and 1.2~meV for the data obtained with the three incident energies, respectively. To map out reciprocal space, the sample was gradually rotated over 100$^{\circ}$ around the [001] crystallographic direction in 0.5$^{\circ}$ steps. The collected data were reduced and analysed using the Horace software~\cite{horace}. A symmetrization procedure was applied during the data reduction, which means that the data from equivalent $\textbf{Q}$-directions in momentum space [for example, $(HH0)$ and $(-H\phantom{|}2H\phantom{|}0)$] were averaged to increase the statistics. Thus, the covered $\textbf{Q}$-space was folded down to a 30$^{\circ}$ sector in the ($HK0$) plane, which is irreducible for the hexagonal system. The TAS measurements were conducted with the fixed final energy $E_{\text{f}}$ = 14.7~meV, PG (002) monochromator and analyzer were used. Two PG filters were put on final energy arm to
remove higher order harmonics. The energy resolution is changing from
1.2 to 2.8 meV from zero to 25 meV energy transfer. All the measurements were performed at a temperature $T$ =  6~K. SpinW software~\cite{spinw} was used to calculate the magnon dispersions within linear spin-wave theory (LSWT).

To evaluate the lattice dynamics, first-principles calculations were carried out using the projector-augmented wave method~\cite{phcalc1} within density functional theory, as implemented in the Vienna \textit{ab initio} simulation package~\cite{phcalc2}. We used the generalized gradient approximation functionals with Perdew-Burke-Ernzerhof parametrization~\cite{phcalc3}. The plane-wave cutoff was set at 600~eV. The size of the k-point mesh for Brillouin zone, based on the Monkhorst-Pack scheme~\cite{phcalc4}, was $7 \times 7 \times 7$. The lattice parameters and ion coordinates were optimized until the residual forces acting on ions became less than 1~meV/\AA. The noncollinear magnetic structure inferred from previous studies was used. The phonons were calculated by constructing a supercell ($2 \times 2 \times 2$) and calculating the force constants using the small-displacement method implemented in PHONOPY~\cite{phcalc5}.

\begin{figure}[t]
        \begin{minipage}{0.99\linewidth}
        \center{\includegraphics[width=1\linewidth]{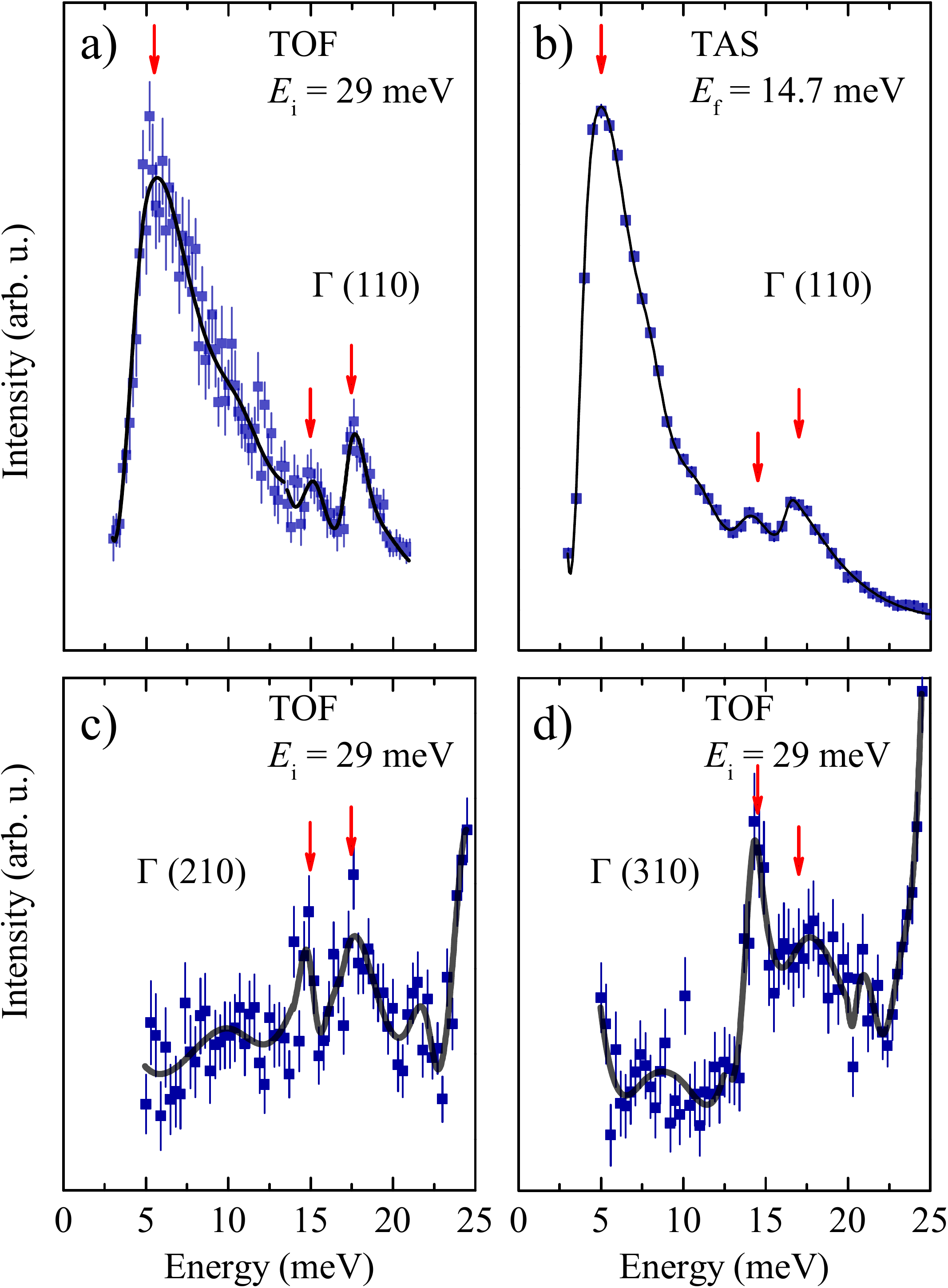}}
        \end{minipage}
        \caption{(color online). Neutron scattering intensity as a function of energy transfer at different constant momentum transfers. (a) The TOF data collected at $\textbf{Q} = (110)$. (b) The TAS measurements at the same momentum. (c)--(d) TOF data at $\textbf{Q} = (210)$ and (310). Error bars represent 1$\sigma$. The incoming or outgoing neutron energy used in different setups is labelled as $E_{\text i}$ or $E_{\text f}$, respectively. Red arrows marks the position of the excitations discussed in the text. Solid lines represent spline interpolations as a guide to the eye.}
        \label{ris:fig1}
\end{figure}

\begin{figure*}
\includegraphics[width=\linewidth]{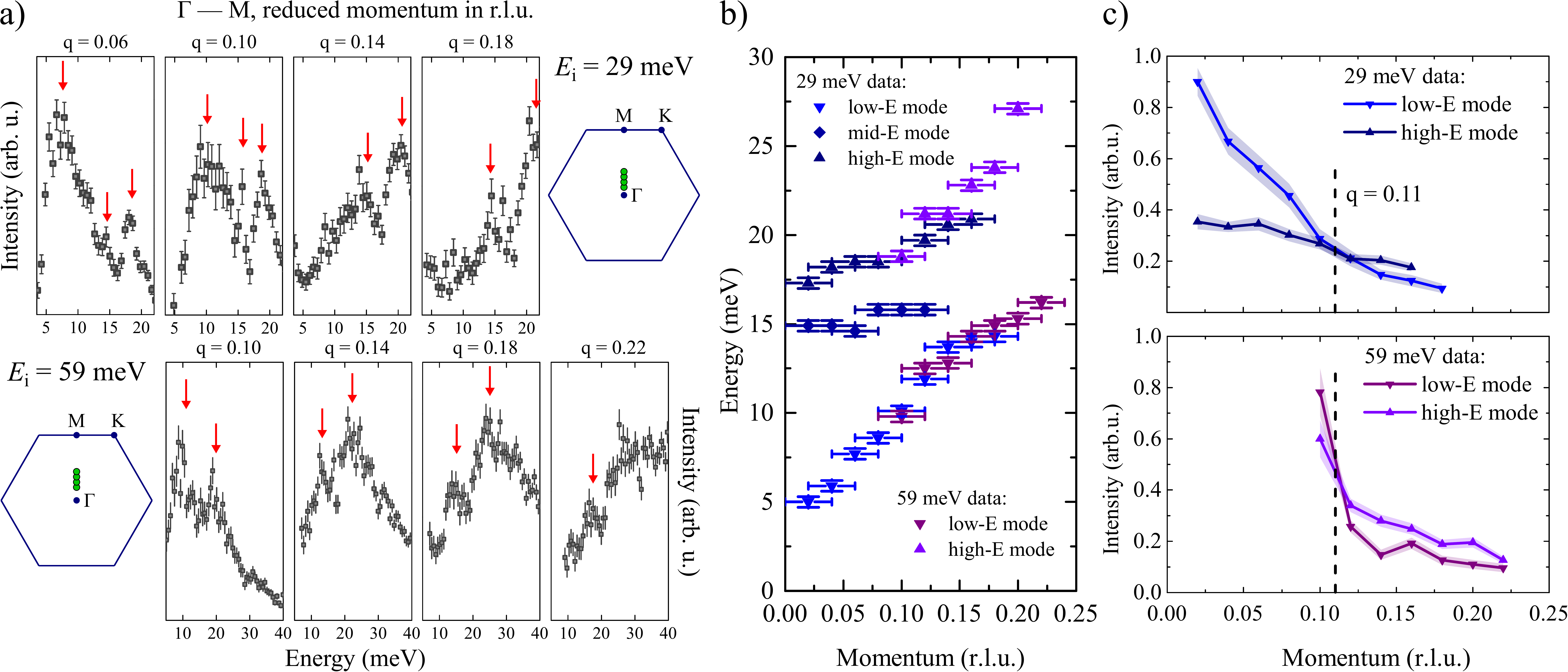}\vspace{3pt}
        \caption{(color online) Low-energy dispersion of the magnetic excitations along the $\mathit{\Gamma}$--$M$ path in the first Brillouin zone (BZ). (a) Constant-$\textbf{Q}$ cuts through the TOF data collected with $E_{\text i} = 29$~meV (top row) and $E_{\text i} = 59$~meV (bottom row). The green dots within the BZ mark the corresponding momenta for which the energy cuts are shown. The integration range along $q$ is $\pm 0.02$~r.l.u. and $\pm 0.05$~r.l.u. for the orthogonal directions. Red arrows mark the positions of the excitations discussed in the text. Error bars represent 1$\sigma$. (b) The peak positions extracted from the energy cuts. The error bars in momentum and energy denote the integration range and the data binning, respectively. (c) Peak intensities of the observed excitations as a function of the momentum. Error bars represent 1$\sigma$.}
        \label{ris:fig2}
\end{figure*}

\section{Results}
\subsection{Excitations at the zone center}

First we consider the intensity of the INS as a function of energy transfer at the zone center $\mathit{\Gamma}(110)$. The (110) reciprocal-space point was previously characterized as the strongest magnetic Bragg peak in the ($HK0$) plane \cite{Sukhanov}. Figure \ref{ris:fig1}(a) and \ref{ris:fig1}(b) show such constant-$\textbf{Q}$ energy scans collected in the TOF and TAS experiments. As can be seen, the two measurements gave identical results. The spectrum is characterized by a very intense peak at $\sim 5$~meV and two additional weaker peaks centred at $\sim 14.5$ and 17~meV. The third feature is approximately two times higher in amplitude than the second 14.5-meV excitation and also 3--9 times lower than the first excitation (seen differently in the TOF and TAS experiments due to the varying orthogonal components of the resolution function). An yearly TAS study of the isostructural compound Mn$_3$Sn reported a similar spectrum at the zone center in the 120$^{\circ}$ magnetic state at 295~K, as well as in the helically modulated AFM phase at 100~K~\cite{old}. The authors of Ref.~\cite{old} also mentioned an observation of three excitations in Mn$_3$Ge at a temperature of 100~K: at 4.5, 14, and 17~meV, which is very close to the results of our measurements. The first peak unambiguously points to gapped spin waves. The two extra excitations may also suggest the presence of second and third spin-wave branches with two different gaps. However, in order to give a possible explanation of the complex splitting of all the spin-wave modes at the $\Gamma$-point, a sophisticated crystal-field scheme is required. The latter is not inherent to 3$d$-metal compounds, where the orbital momentum is usually quenched.

A more plausible explanation can be given if spectra at $\mathit{\Gamma}(110)$ and $\mathit{\Gamma}(210)$ are compared [Fig. \ref{ris:fig1}(c)]. At (210), there is only a weak magnetic reflection due to both the structure factor of the coplanar triangular spin configuration and the reduced magnetic form-factor~\cite{Sukhanov}. Correspondingly, the magnons in the vicinity of (210) carry only vanishingly small spectral weight. Therefore, all the magnetic excitations should either be absent or show very small intensities. As expected, the 5-meV peak is suppressed. On the contrary, the 14.5- and 17-meV excitations remain visible, yet with smaller intensities as can be noticed by the weakened signal-to-noise ratio. The fact that these two extra excitations appear in both the (110) and (210) zone centres indicates the phonon nature of the excitations. The spectrum collected at $\textbf{Q} = (310)$, which is larger in absolute value than (210) [Fig. \ref{ris:fig1}(d)] shares the same characteristics. The 5-meV magnon is absent and there are two distinguishable excitations at exactly same positions as in Fig.~\ref{ris:fig1}(a)--(c). One can also observe a high-intensity phonon located at $\sim 25$~meV and a weaker one at $\sim 22$~meV in both $\mathit{\Gamma}(210)$ and $\mathit{\Gamma}(310)$ energy cuts. Importantly,  the lattice dynamics contributes to the INS cross-section with a prefactor proportional to $Q^2$, which increases with higher $H$ in $(H10)$ for the discussed reciprocal-space points. As follows from this, pure optical phonons are expected to show weaker intensities at (110), whereas they are significantly enhanced at (110) as evidenced by large intensity of the excitations. The latter points to a mixed magnon-phonon character of the 14.5- and 17-meV features.

Such an extra excitation observed by INS in the cycloidal magnet TbMnO$_3$ was interpreted as a magnon-phonon mode in Ref.~\cite{ref_gamma_1}. Multiple resonances were resolved in a study of the frustrated magnet MgCr$_2$O$_4$, where they were interpreted as the molecular excitations of the spin clusters~\cite{ref_gamma_2}. Later, the magnetoelastic origin of those excitations was also suggested as a possible explanation~\cite{ref9}. Comparing the INS intensities at high and low momenta, the authors of Ref.~\cite{ref_gamma_3} proposed the existence of the hybrid excitation described as a crystal-field level coupled to a transverse acoustic phonon in the pyrochlore compound Tb$_2$Ti$_2$O$_7$.

\subsection{Low-energy dispersion}

\begin{figure*}
\includegraphics[width=\linewidth]{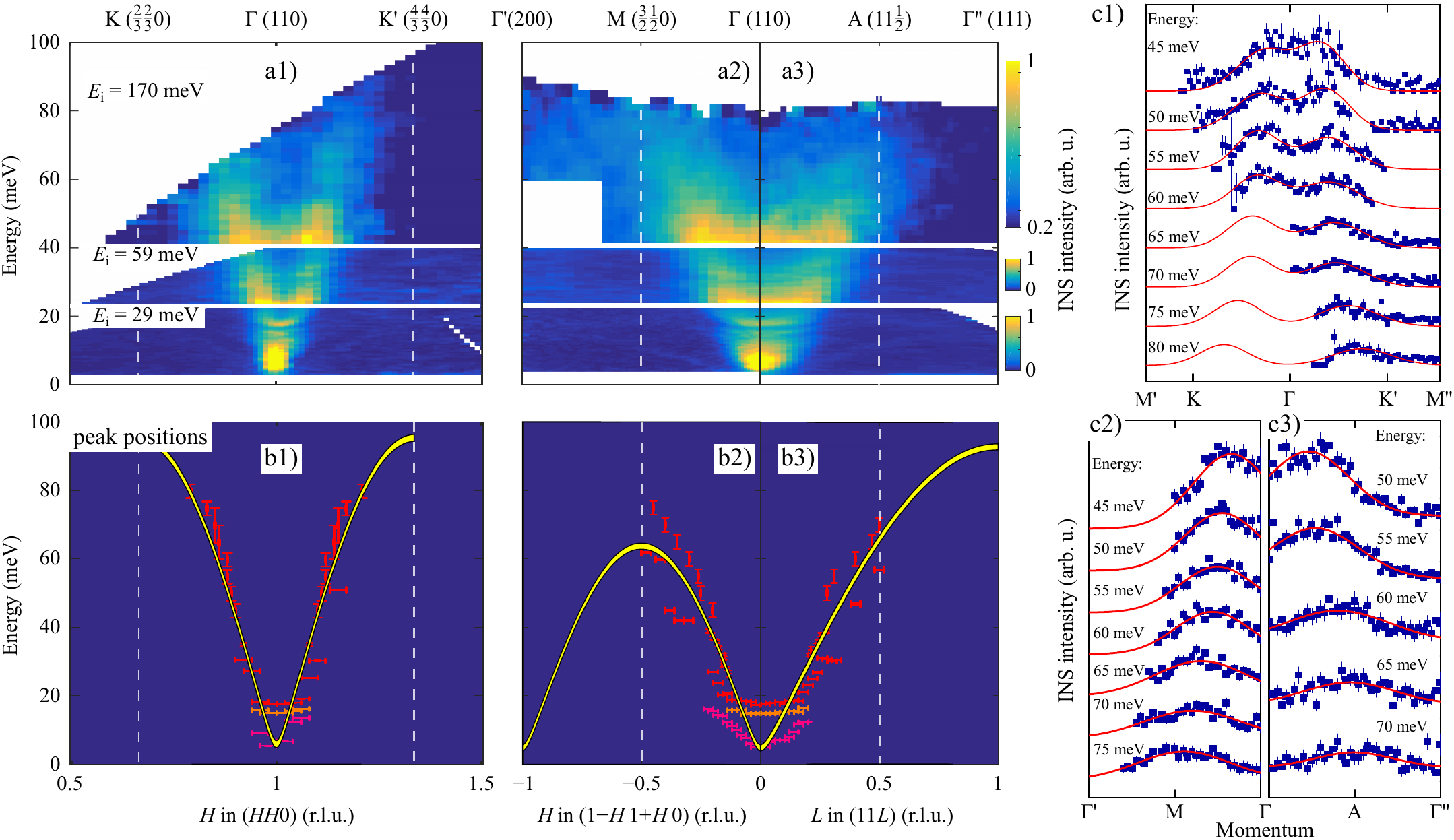}\vspace{3pt}
        \caption{(color online). Momentum-energy cuts through the TOF data and results of the fitting. (a1)--(a3) Slices of data collected with different $E_{\text{i}}$ are represented in different energy bands as indicated in the left-hand side of panel (a1). The energy is shown as a function of momenta for high-symmetry directions of the BZ as specified above the panels (a1)--(a3). The momentum integration range in directions orthogonal to the image for $E_{\text{i}}=170$ and for $E_{\text{i}}=59, 29$ meV was respectively set (in r.l.u.): (a1) $\pm 0.1, \pm 0.15; \pm 0.1, \pm 0.1$; (a2) $\pm 0.1, \pm 0.15; \pm 0.05, \pm 0.1$; (a3) $\pm 0.1, \pm 0.1; \pm 0.05, \pm 0.1$. The color maps at different energy windows are not to scale. (b1)--(b3) Extracted INS peak positions as obtained from constant-energy and constant-momentum cuts through the experimental data. Horizontal (vertical) error bars denote the integration range in momentum (energy transfer). Solid yellow lines are calculated spin-wave dispersions in the simplified model, as described in the text. (c1)--(c3) Constant-$E$ cuts through the TOF data with $E_{\text i} = 170$~meV, solid lines are Gaussian fits.}
        \label{ris:fig3}
\end{figure*}

Having determined the distribution of the INS intensity at the $\mathit{\Gamma}$-point, we can focus on following the dispersion of the observed excitations. There are three high-symmetry paths in the Brillouin zone (BZ) of a hexagonal lattice: $\mathit{\Gamma}$--$M$ (linking the zone center with an edge of the hexagon), $\mathit{\Gamma}$--$K$ (a vertex of the hexagon), and $\mathit{\Gamma}$--$A$ (perpendicular to the hexagonal plane). First, we demonstrate ф detailed analysis of the low-energy sector (with respect to the top of the band) of the magnetic excitations along the $\mathit{\Gamma}$--$K$ path based on the energy cuts through TOF data at different momenta between the (110) and (200) reciprocal-space points. Figure~\ref{ris:fig2}(a) shows the INS intensity extracted from the measurements with $E_{\text i} = 29$~meV and $E_{\text i} = 59$~meV. The latter results in a wider energy-transfer range but lower resolution. The two datasets represent the collected statistics within the energy window from 3.5 to 22~meV and from 5 to 40~meV, respectively. The covered energy transfer allows one to track all three branches of the magnetic excitations from the zone center to the momenta that are in the vicinity of the middle point between $\mathit{\Gamma}$ and $M$. The low-lying magnon with an energy of $\sim 5$~meV at $q = 0$ [Fig.~\ref{ris:fig1}(a),(b)] is observed at $\sim 7.5$~meV at $q = 0.06$, where $q$ denotes the reduced momentum along $\mathit{\Gamma}$--$M$ (with $q = 0.5$ at the zone boundary). On the contrary, the second and third excitations, identified as hybridized magnon-phonon states, remain at approximately the same energies. As further evidenced by the data at $q = 0.1$ [Fig.~\ref{ris:fig2}(a)], the magnon-phonon branches show a weak upward dispersion, whereas the lower spin-wave reaches the energy of $\sim 10$~meV, thus exhibiting a linear dispersion. The second mode, being the weakest in intensity, disappears or merges with the first (low-lying) mode at $q = 0.14$. In accordance with that, only two branches are observed at higher $q = 0.18$ in the $E_{\text i} = 29$~meV dataset. The middle mode is unresolved in the energy cuts from the $E_{\text i} = 59$~meV data, where only the low- and high-energy branches are clearly visible. The peak positions of these two excitations inferred from the different datasets are in full agreement as can be compared for $q = 0.1, 0.14,$ and 0.18 in Fig.~\ref{ris:fig2}(a). The intense peak from the high-energy branch can be well resolved at $q = 0.18$, whereas it transforms to a broad hump at $q = 0.22$. In contrast, the low-energy branch can still be identified as a sharp feature in the energy cut at $q = 0.22$, but it vanishes completely at higher momentum transfers.

The peak positions extracted from the constant-$\textbf{Q}$ cuts of the TOF data along the $\mathit{\Gamma}$--$M$ path are summarized in Fig.~\ref{ris:fig2}(b). As was previously mentioned, the 5-meV-gapped mode demonstrates a linear dispersion at momenta close to the center of BZ. The slope of the mode remains constant until $q \sim 0.11$, after which the slope noticeably decreases. Interestingly, the dispersion of the third branch experiences a significant upward change at the same momentum. The latter is a characteristic of level repulsion. A similar anticrossing point at the reduced momentum $q \simeq 0.185$ was found in YMnO$_3$~\cite{ref11,ref14}. The signatures of mode anticrossing are also found in the $Q$-dependence of the observed peak intensities [Fig.~\ref{ris:fig2}(c)]. The first branch has two times higher intensity in the vicinity of the $\mathit{\Gamma}$-point. The difference in the intensities of the first and the third branches becomes smaller at increasing momenta $0 < q < 0.11$. At the reduced momentum $q = 0.11$, the two excitations show approximately the same intensity. Furthermore, at higher momenta the third mode acquires higher intensity than that of the first mode. The same crossover in the relative intensities of the observed modes can be seen in both $E_{\text i} = 29$ and $E_{\text i} = 59$~meV datasets. The collinear AFM MnWO$_4$ was reported to demonstrate the same rapid drop in the INS intensity of the two low-energy magnetoelastic modes, which were observed only within the momentum of 0.05 r.l.u away from the magnetic zone center~\cite{ref_disp_1}.

\subsection{Magnetic excitations across the entire BZ}

\begin{figure}[t]
        \begin{minipage}{0.99\linewidth}
        \center{\includegraphics[width=1\linewidth]{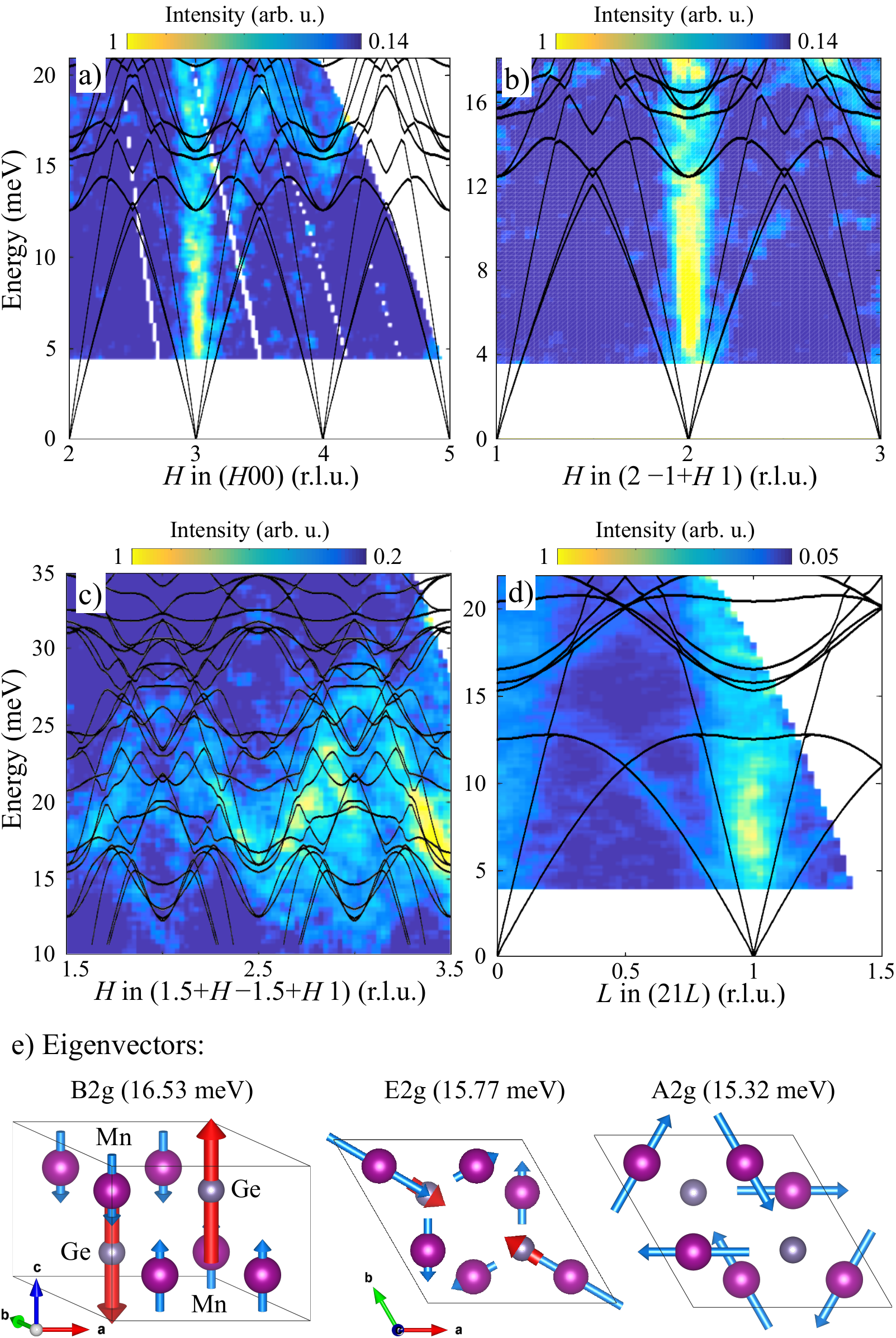}}
        \end{minipage}
        \caption{(color online). Measured phonon dispersions and results of the \textit{ab initio} calculations of the lattice dynamics. (a)--(d) Energy-momentum cuts through the TOF data collected with $E_{\text i} = 29$ and 59~meV, calculated phonon bands (black lines) are plotted over the experimental data for comparison. (e) Eigenvectors of some of the optical phonon modes at the $\mathit{\Gamma}$-point, the arrows depict the atomic displacement; the magnetic structure of Mn atoms is not shown.}
        \label{ris:fig4}
\end{figure}

Next we turn to the overview of the spin-wave dispersion across the entire BZ inferred from the energy-momentum cuts through the TOF data, which are demonstrated in Figs. \ref{ris:fig3}(a1)--\ref{ris:fig3}(a3) for the high-symmetry paths $K$--$\mathit{\Gamma}$--$K^{\prime}$, $\mathit{\Gamma^{\prime}}$--$M$--$\mathit{\Gamma}$, and $\mathit{\Gamma}$--$A$--$\mathit{\Gamma^{\prime\prime}}$, respectively. The datasets collected with different $E_{\text{i}}$ are represented separately in different energy windows to combine a broad accessed energy range at $E_{\text{i}}=170$~meV and a good energy resolution at lower incident neutron energies $E_{\text{i}}=59$ and 29~meV. One can see an intense spot located at the $\mathit{\Gamma}$-point at an energy of 5~meV in accordance with the TAS results [Fig.~\ref{ris:fig1}(b)]. The spot transforms into a narrow cone of intensity towards higher energies up to $\sim$~14.5~meV where the second excitation emerges. The observed intensity in the (5.0--14.5)~meV range can be interpreted as a steep spin-wave mode with anisotropic stiffness. The excitation at 14.5~meV does not result in any visible excitation away from the $\mathit{\Gamma}$-point, whereas the high-intensity point at 17~meV forms the second dispersive mode with slightly softer slope. Only one magnon branch is resolved at higher energies and seen as a solitary expanding cone of intensity. The dispersive excitation already reaches an energy as high as $\sim 85$~meV at the reduced wave-vector $q \sim (0.19, 0.19, 0)$ r.l.u. in the $\mathit{\Gamma}$--$K$ direction. The spin waves show a softer dispersion along the $\mathit{\Gamma}$--$M$ path and acquire an energy of $\approx 65$~meV in the vicinity of the $M$-point. The mode demonstrates signatures of a downturn at $M$, however the measured spectral weight significantly reduces at momenta along $M$--$\mathit{\Gamma^{\prime}}$ and no replica can be seen. A somewhat softer dispersion is observed along the $\mathit{\Gamma}$--$A$ path for the energies up to $\approx 40$--$50$~meV where the excitations become more isotropic with respect to the plane defined by the $\mathit{\Gamma}$-, $M$-, and $A$-points in the high-energy region. The INS intensity in the vicinity of the $A$-point suggests a collective excitation with the energy of $\sim 65$~meV. No intensity is further observed at momenta between the $A$- and $\mathit{\Gamma^{\prime}}$-points. Our results on the low-energy part of the spectrum agree with the yearly TAS study of Mn$_3$Ge~\cite{old}, where the excitations along $(HH0)$ and $(00L)$ were probed up to the energy transfer of 25~meV.

\begin{table}[b]
\small\addtolength{\tabcolsep}{+7pt}
\caption{Exchange parameters (simplified model of independent AFM spin chains).}
\label{tab:tab1}
 \begin{tabular}{c c c c}
 \hline\hline
 & $\mathit{\Gamma}$--$K$ & $\mathit{\Gamma}$--$M$ & $\mathit{\Gamma}$--$A$--$\mathit{\Gamma}$ \\ [1ex] 
 \hline
 $JS$ (meV) & $47.8 \pm 5.9$ & $31.8 \pm 1.8$ & $46.4 \pm 3.5$\\ [1ex]   
 \hline\hline
\end{tabular}
\end{table}

The complex low-energy dynamics, which was presented for the $\mathit{\Gamma}$--$M$ path in details in Fig.~\ref{ris:fig2}, takes place for all three high-symmetry directions in the reciprocal space. We conducted the same analysis of the low-energy ($<30$~meV) data for the $\mathit{\Gamma}$--$K$ and the $\mathit{\Gamma}$--$A$ paths. For this, we used constant-$\textbf{Q}$ energy cuts to extract positions of the relatively sharp INS peaks from three modes. The low-energy mode vanishes at momentum $\sim (1$~1~0.2)~r.l.u. in the $(11L)$ direction. This is comparable with the reduced momentum at which the same mode loses its spectral weight along (1$-H$~1$+H$~0). On the contrary, the dispersion of the first mode is observed only up to $q \sim 0.06$ along $\mathit{\Gamma}$--$K$. In order to track the dispersion of the high-energy mode, we considered constant-energy momentum cuts through the TOF data collected with $E_{\text i} = 59$~meV for the energy window (30--40)~meV and with $E_{\text i} = 170$~meV for the energy transfers above 40~meV. In addition, the peak positions in the energy cuts from the 170-meV data were extracted at momenta close to the zone boundary. The resultant energy-momentum positions of the magnetic excitations are summarized in Figs.~\ref{ris:fig3}(b1)--\ref{ris:fig3}(b3). Figures~\ref{ris:fig3}(c1)--\ref{ris:fig3}(c3) show Gaussian fits to the broad INS peaks in the constant-$E$ cuts of the 170-meV dataset. The peak positions extracted from the $E$- and $Q$- cuts gave slightly different results due to the nonfocusing conditions of the instrumental resolution function and rapidly decreasing spectral weight of the observed excitations at high energies.

In an attempt to quantify the high-energy part of the magnon dispersion, i.e. away from the part of the spectrum dominated by magnon-polaron excitations, we employed the simple model of an AFM spin chain. For this purpose, the experimental spin-wave dispersions along $\mathit{\Gamma}$--$K$, $\mathit{\Gamma}$--$M$, and $\mathit{\Gamma}$--$A$ were treated independently. The model assumes only one Heisenberg exchange parameter between the nearest-neighbouring spins of the chain. The spin-wave dispersion of the AFM chain was calculated within the linear spin-wave theory (LSWT) (implemented in the SpinW software package~\cite{spinw}) and fitted to the extracted peak positions for the energies above 25~meV separately for $\mathit{\Gamma}$--$K$, $\mathit{\Gamma}$--$M$, and $\mathit{\Gamma}$--$A$ [Fig.~\ref{ris:fig3}(b1)--\ref{ris:fig3}(b3)]. To fit the experimental dispersion along $(11L)$, the BZ of the spin chain was doubled. The resulting exchange parameters found in the frames of this simple model are listed in Table~\ref{tab:tab1}. A realistic spin model that can consistently account for the full spin-wave spectrum is out of the scope of the present study and should be addressed in future theoretical works.

It is worth to briefly discuss the spin-wave models that were proposed in the previous studies of Mn$_3$Sn~\cite{old,new} and Mn$_3$Ge~\cite{old}. The authors of Ref.~\cite{old} suggested a magnetic Hamiltonian for Mn$_3$Ge that takes into account five exchange interactions (two intralayer and three interlayer interactions). Based on their limited experimental data, Cable \textit{et. al.} extracted the values of all the assumed exchange interactions that were quantified as follows. The nearest-neighbor (interlayer) and next nearest-neighbor (intralayer) interactions, as well as the interaction via two layers, are antiferromagnetic, whereas the interactions at higher distances (in-plane, between the spins from the same sublattice, and out-of-plane, between the spins from different sublattices) are ferromagnetic. When tried to apply this model to our data, we found out that the parameters, reported by Cable \textit{et. al.}, do not reproduce the correct magnetic ground state. If the balance between the intralayer and interlayer interactions are modified by $\sim 20$\%, the correct ground state can be achieved. Whilst the modified exchange scheme can show some distant similarity to the experimentally observed low-energy part of the excitations, it fails completely at higher energies.

In a recent study~\cite{new}, Park \textit{et. al.} used a model that includes exchange interactions up to the 8th coordination sphere to describe the INS spectra of Mn$_3$Sn. This model can be considered as an extended version of the model proposed in Ref.~\cite{old}. The validity of their findings is questionable, as they constructed a complex 3D model by measurements of the excitations only in the $(HK0)$ plane. The model proposed in Ref.~\cite{new} provides a limited agreement with the magnon spectrum of Mn$_3$Sn and does not seem to be applicable to Mn$_3$Ge. The main discrepancy of the models proposed in Refs.~\cite{old}~and~\cite{new} originates in the fact that two distinct (with an energy difference well above the instrumental resolution and approximately equal spectral weight) spin-wave modes are predicted above $E \sim 40$~meV, but only one is observed. Future studies may help clarify this discrepancy.

\subsection{Phonon dynamics in the absence of magnon-phonon coupling}

\begin{table}
\small\addtolength{\tabcolsep}{+2pt}
\caption{Energies, symmetries and eigenvectors $(x,y,z)$ of the low-lying optical phonons at the $\mathit{\Gamma}$-point obtained from \textit{ab initio} calculations.}
\label{tab:tab2}
 \begin{tabular}{c c c c}
 \hline\hline
 $E$ (meV) & Symmetry & Eigenvectors Mn & Eigenvectors Ge \\ [1ex] 
 \hline
 12.52 & E2g & $(-0.19, -0.15, 0)$; & $(-0.47, -0.03, 0)$; \\ [1ex]

  &  &  $(-0.18, \phantom{-}0.11, 0)$;  &  \\ [1ex]

  &  & $(-0.41, -0.01, 0)$; &  \\ [1ex] \hline

 15.32 & A2g & $(\phantom{-}0.20, \phantom{-}0.35, 0)$; & $(0, 0, 0)$; \\ [1ex]

  &  &  $(\phantom{-}0.20, -0.35, 0)$;  &  \\ [1ex]

  &  & $(-0.41, \phantom{-0.0}0, 0)$; &  \\ [1ex] \hline

 15.77 & E2g & $(\phantom{-}0.48, -0.28, 0)$; & $(-0.17, 0.11, 0)$; \\ [1ex]

  &  &  $(\phantom{-}0.22, \phantom{-}0.12, 0)$;  &  \\ [1ex]

  &  & $(\phantom{-0.0}0, -0.30, 0)$; &  \\ [1ex] \hline

 16.53 & B2g & $(0, 0, 0.25)$; & $(-0.56, 0, 0)$; \\ [1ex]

  &  &  $(0, 0, 0.25)$;  &  \\ [1ex]

  &  & $(0, 0, 0.25)$; &  \\ [1ex] \hline

 20.43 & B2u & $(-0.20, -0.35, 0)$; & $(-0.56, 0, 0)$; \\ [1ex]

  &  &  $(-0.20, \phantom{-}0.35, 0)$;  &  \\ [1ex]

  &  & $(\phantom{-}0.41, \phantom{-0.0}0, 0)$; &  \\ [1ex]

 \hline\hline
\end{tabular}
\end{table}

To further elucidate the nature of the complex excited states observed in the vicinity of the $\mathit{\Gamma}$-point and the energy range of $\sim 10$--25~meV, we carried out lattice-dynamics calculations. Because a large part of reciprocal space was covered in the TOF experiment, the data obtained for high momentum transfers can be used to differentiate the phonon modes from the magnons and the magnon polarons, carrying the majority of the spectral weight at low momentum transfers. We found that the calculated phonon bandwidth is in a very good agreement with the experiment when a small renormalizing coefficient $\alpha = 1.08$ is applied to the calculated phonon frequencies. The renormalized calculated phonon branches are demonstrated in Figs.~\ref{ris:fig4}(a)--\ref{ris:fig4}(d) along with the experimental data presented for a number of high-symmetry momentum directions. As can be seen, the calculated phonon dispersions do not deviate by more than 10\% from the experimental spectra. The main discrepancy between the theoretical phonon dispersions and the data can be noticed in the low-lying optical phonons in the vicinity of the zone center. This is expected in the system with a magnon-phonon interaction, as this interaction is not taken into account in the present calculations. The intense steep dispersion stemming from the (300) and (211) reciprocal-space points in Fig.~\ref{ris:fig4}(a) and  Figs.~\ref{ris:fig4}(b), \ref{ris:fig4}(d) are weaker replicas of the (110) magnetic excitations. As one can see, the calculations reproduce the longitudinal acoustic phonon branch observed in $(H00)$ for $H$ from 3 to 3.5 r.l.u. well~[Fig.~\ref{ris:fig4}(a)]. The transverse acoustic phonons have no spectral weight along $(H00)$ due to the polarization factor, but can be seen in (2~$-$1$+H$~1)~[Fig.~\ref{ris:fig4}(b)], also in good agreement with the calculated dispersion. The transverse acoustic mode dispersing along the $(00L)$ direction can be seen in Fig.~\ref{ris:fig4}(d). The low-lying optical modes ($\sim 15$--20~meV) that stem from (400), (221), (210), and (211) are experimentally resolved and demonstrate a slight deviation (by less than 10\%) with the calculation close to the zone center, whereas show an excellent agreement at the zone boundaries for both in-plane [Figs.~\ref{ris:fig4}(a), \ref{ris:fig4}(b)] and out-of-plane reduced momenta [Fig.~\ref{ris:fig4}(d)]. The phonon branches at higher energies are presented in Fig.~\ref{ris:fig4}(c). The signal-to-noise ratio allows one to identify the phonon states with the energies as high as $\sim 32$~meV at the $\mathit{\Gamma}$-point, $\sim 30$~meV at the $M$-, and $\sim 27$~meV at the $K$-points.

The overall success of the phonon calculations in the application to the experimental data allows us to identify all the lattice vibrations in the zone center, where the strong magnon-polaron excitations were observed. The details on the first five phonon states with the reduced wave-vector $q = 0$ are listed in Table~\ref{tab:tab2}. The eigenvectors of three of them are also illustrated in Fig.~\ref{ris:fig4}(e). As seen in Table~\ref{tab:tab2}, the A2g mode only involves the displacement of Mn atoms, whereas all the other phonons are characterized by non-zero eigenvectors of both Mn and Ge atoms. Four of the five shown lattice excitations are characterized by the in-plane movement of the ions and only the 16.53-meV level describes the out-of-plane oscillations.

\section{Discussion and conclusions} 

The optical phonon bands listed in Table~\ref{tab:tab2} are considered as primary candidates for the observed hybridization with the magnon mode. Because of the high $T_{\text N}$ of 370~K, the magnon bandwidth is almost three times greater than the top energies of the lattice vibrations ($\simeq 90$~ meV vs $\simeq 35$~ meV). Due to this extreme stiffness of the spin-wave mode in comparison with the phonon dispersions, the intersection of the magnon dispersion with the first optical phonons occurs at a small momentum. Thus, the resulting magnon-polaron excitations are seen in the vicinity of the BZ center. Seemingly, one can associate the observed first (14.5~meV) magnon-polaron state as the result of the hybridization with the lowest optical phonon with the calculated bare energy of 12.52~meV, which implies the $\simeq 16$~\% renormalization of the bound state. The next three phonon bands have similar energies with less than 1.5-meV difference at the  $\mathit{\Gamma}$-point. These lattice vibrations may contribute to the observed second magnon-polaron state at 17~meV. Therefore, the 17-meV excitation may actually have an unresolved fine structure caused by the hybridization of each of these three optical phonons with the spin waves. In addition, magnetoelastic coupling may split the doubly-degenerate phonon modes as was shown for the frustrated spinel ZnCr$_2$O$_4$~\cite{disc_extra1,disc_extra2,disc_extra3} The fourth phonon level is separated by an energy of $\simeq 4$~meV at the $\mathit{\Gamma}$-point and is too high to contribute to the second bound excitation. We note, that only excitations observed in the vicinity of the magnon-phonon anticrossing point (from the $\mathit{\Gamma}$-point to $q \sim 0.1$~r.l.u. for the two upper branches, and from $q \sim 0.1$ to $\sim 0.2$ r.l.u. for the lower branch) can be qualified as the magnon polarons.

The impact of the magnetoelastic coupling on the spectrum of Mn$_3$Ge seems to be in a distinct contrast to what was observed in some other AFMs with triangular spin structures. In LiCrO$_2$~\cite{ref9}, the magnon-phonon interaction caused a significant downward renormalization of the bare magnon dispersion at the $M$-point of the hexagonal BZ, a so-called roton-like minimum~\cite{disc2}.The same scenario realized in CuCrO$_2$~\cite{ref17}. No such minima are observed in the present study of Mn$_3$Ge. There is also no additional spectral weight at the intermediate energy optical phonons in the vicinity of the $M$-point (at low momenta within the range of the magnetic form-factor), as was reported for (Y,Lu)MnO$_3$~\cite{ref10} and CuCrO$_2$~\cite{ref17}. The latter may be due to the significantly different band widths of the magnon and the phonons in Mn$_3$Ge. Dispite this, YMnO$_3$ and Mn$_3$Ge demonstrate similar magnon-phonon anticrossings close to the BZ center~\cite{ref11,ref13,ref14}. It is also important to note that it is the acoustic phonon mode that was observed to hybridize with the spin-wave excitations in the abovementioned noncollinear AFMs~\cite{ref9,ref11,ref13,ref14,ref17} and in the collinear ferrimagnet YIG~\cite{ref2,ref3,ref4,ref7,ref_extra_1}. Therefore, Mn$_3$Ge represents a rather unique example of a system where the magnon-polarons originate from the coupling with the optical phonons.

In a recent INS study, Park \textit{et al}.~\cite{new} presented a spectrum of magnetic excitations in the isostructural noncollinear AFM Mn$_3$Sn across the entire BZ. The obtained spectrum was analyzed using a model, developed in the previous study~\cite{old}, and was claimed to be strongly affected by a magnon damping effect. It is not excluded that the discussed in Ref.~\cite{new} magnon damping might be related to magnon-phonon interaction in Mn$_3$Sn, which has not been so far discussed for this compound.

To conclude, we have conducted neutron spectroscopy measurements in which we have covered a large part of 4D ($E$,\textbf{Q}) reciprocal space. The collected data reveals a gapped spin-wave mode that stems from the center of the crystallographic BZ, which is a characteristic of an AFM structure with the $\textbf{k} = 0$ propagation vector. The spin-wave mode is found to induce anticrossing  points with the low-lying optical phonons in the vicinity of the zone center. The resulting magnon-polaron excitations carry the spectral weight at 14.5 and 17 meV energy transfer at the $\mathit{\Gamma}$-point. They demonstrate a weak dispersion before reaching the anticrossing point, located at the reduced momenta $\sim$ 0.1--0.15 r.l.u (depending on the reciprocal-space direction), after which they eventually vanish. Furthermore, the first-principle lattice-dynamics calculations showed a good agreement with the experimentally obtained phonon spectra and allowed the confirmation of the hybrid magnetoelastic nature of the discussed excitations.

The magnetic excitations with momenta beyond the magnon-polaron anticrossing point are described by a steeply dispersing anisotropic spin-wave mode that reaches energies of $\sim 95$~meV at the BZ boundary at the $K$-point and  $\sim 65$~meV at the $M$- and $A$-points. The spin-wave dispersion along each high-symmetry direction of the BZ can be well described by the simple model of the Heisenberg AFM spin chain. However, the spin Hamiltonian that can simultaneously account for the whole magnon spectrum remains to be constructed. Thus, the spin-wave dispersion across the entire BZ presented in this study can be used for future tests of advanced spin models.

\section*{Acknowledgments}
We thank S. E. Nikitin for many stimulating discussions. A.S.S. acknowledges support from the International Max Planck Research School for Chemistry and Physics of Quantum Materials (IMPRS-CPQM). The work at the TU Dresden was funded by the German Research Foundation (DFG) in the framework of the Collaborative Research Center SFB 1143 (project C03), Würzburg-Dresden Cluster of Excellence \textit{ct.qmat} (EXC 2147, project-id
39085490), and the Priority Program SPP 2137 ``Skyrmionics''.

\end{document}